\documentclass[preprint2]{aastex}
\newcommand{\mdot}{$\dot{M}$}
\newcommand{\maccdot}{$\dot{M}_{\rm acc}$}
\newcommand{\leps}{{\large $\epsilon$}}
\newcommand{\etal}{et\,al.}
\shorttitle{SU~UMa-type Cataclysmic Variables}
\shortauthors{Coel Hellier}
\begin{document}
\title{On Echo Outbursts and ER~UMa Supercycles in SU~UMa-type
Cataclysmic Variables}
\author{Coel Hellier}
\affil{Department of Physics, Keele University, Staffordshire ST5 5BG, U.K.}
\email{ch@astro.keele.ac.uk}
\begin{abstract}
I present a variation on Osaki's tidal-thermal-instability
model for SU~UMa behavior. I suggest that in systems with the lowest
mass ratios, the angular-momentum dissipation in an eccentric disk
is unable to sustain the disk on the hot side of the thermal
instability. This decoupling of the tidal and thermal instabilities
in systems with $q$\,$\la$\,0.07 allows a better explanation of the 
`echo' outbursts of EG~Cnc and the short supercycles of RZ~LMi and DI~UMa. 
The idea might also apply to the soft X-ray transients.
\end{abstract}


\section{Introduction}
The SU~UMa subclass of cataclysmic variable is characterised
by two distinct accretion-disk instabilities. The first is
a thermal instability caused by partially ionized hydrogen, which
results in dwarf-nova outbursts (e.g.\ Cannizzo 1993).
The second is a tidal instability
occurring when orbits in the disk resonate with the
secondary-star orbit, driving the disk elliptical and causing
it to precess with a period of a few days (e.g.\ Whitehurst \&\ 
King 1991). The 3:1 resonance that is 
responsible occurs at a disk radius of $\approx$\,0.46$a$, where
$a$ is the binary separation. In the elliptical disk the tidal 
stresses vary with the interaction of the orbital and
precessional periods, to produce excess light recurring with
a period a few percent longer than the orbital period,
called a superhump.

\section{SU~UMa supercycles}
Observationally, SU~UMa stars show a succession
of normal outbursts, caused by the thermal instability alone,
followed by a longer-lasting superoutburst, during which the
presence of superhumps implies that the tidal
instability is also excited. This pattern (called a supercycle)
then repeats.

Osaki (1989; 1996) has presented a model (the tidal-thermal-instability 
or TTI model) which combines the instabilities to
explain the supercycle. He suggests that a normal outburst
depletes the disk by less than the matter accumulated since the
last outburst.  The disk therefore grows over a succession of
outbursts until, at the onset of another outburst, it expands
beyond 0.46$a$ and becomes tidally
unstable. The tidal dissipation of the eccentric state enhances
the mass-flow through the disk, sustaining the disk in its
hot state for a longer superoutburst,
and so draining the disk of the majority of its
material. Once the disk has shrunk to $\approx$\,0.35$a$, neither
the eccentricity nor the hot state can be sustained, and the
disk reverts to a cold, circular quiescence, to begin the cycle
anew.

The TTI model is successful in explaining typical SU~UMa stars,
which have supercycles of 100--1000 d. However,
some SU~UMa stars (those with abnormally short or abnormally long
supercycles) have outburst patterns that don't fit the standard
model. This has led Hameury, Lasota \& Warner (2000) to present
alternative models involving 
evaporation of the inner disk and irradiation-induced
increases in the mass-transfer rate. The purpose of this paper is
to suggest that the TTI model can, after all, explain 
the anomalous systems, given one alteration.

\begin{figure}[t]   % Fig 1
\plotone{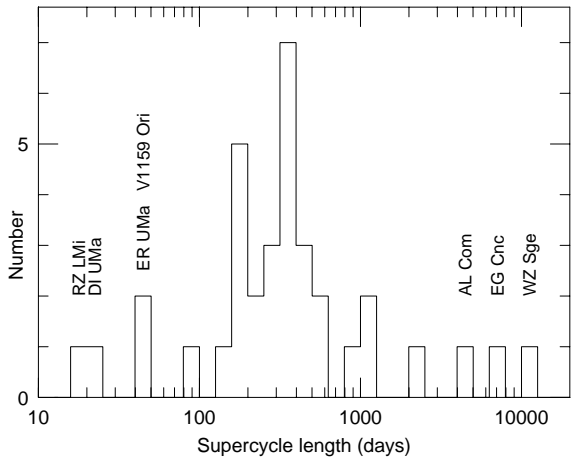}
\caption{The supercycle lengths of SU~UMa stars. Those with abnormally
long supercycles are the WZ~Sge stars, while those with abnormally short
supercycles are the ER~UMa stars.  Data from Warner (1995), supplemented
by papers mentioned in the text.} 
\end{figure}

\subsection{The problem of `echo' outbursts in WZ~Sge systems}
The term `WZ~Sge star' is a shorthand for SU~UMa systems with
supercycles lasting decades (see Fig.~1). Part of the explanation
for this is that the mass-transfer rates, \mdot, are very low, typically
10$^{15}$ g s$^{-1}$, compared to 10$^{16}$ g s$^{-1}$ for
normal SU~UMa stars (Osaki 1995a). Also, such stars rarely or
never show normal outbursts between the superoutbursts.  This
implies that material does not accumulate in
the inner disk, which requires an abnormally low
viscosity, inner-disk evaporation, or a magnetic propeller
(e.g.\ Osaki 1995a; Hameury \etal\ 2000; Wynn, Leach \& King 2000).
However, I do not address this issue here.

Instead, my concern is the `echo' outbursts seen following
superoutbursts in WZ~Sge stars.  These were most striking in
EG~Cnc, which, after the first superoutburst for 19 years,
showed a rapid succession of 6 shorter outbursts, spanning one 
month, before dropping by a further magnitude to full
quiescence (Fig.~2; Patterson \etal\ 1998).      Similar, though
less pronounced, echos have been seen in other WZ~Sge systems
[e.g.\ AL~Com (Patterson \etal\ 1996), UZ~Boo (Kuulkers, Howell \&\
van Paradijs 1996), and WZ~Sge itself (Patterson \etal\ 1981)].

The first problem raised by EG~Cnc is that superhumps were
seen throughout the period of echo outbursts, even during
quiescence (Patterson \etal\ 1998). This contradicts the
standard TTI model, in which the superoutburst ends with both
the thermal and tidal instabilities shutting down.

The second problem is that the mean {\it accretion\/} rate
is clearly higher over the echo period than in full quiescence.
In the TTI model this requires an ad-hoc increase of the
quiescent viscosity to an alpha parameter, $\alpha_{\rm cold}$,
of 0.1 (Osaki, Shimuza \&\ Tsugawa 1997), a value more typical
of a disk in outburst. Alternatively, Hameury \etal\ (2000)
have suggested that irradiation of the disk by a hot white
dwarf produces the echos.

\begin{figure}[t]   % Fig 2
\plotone{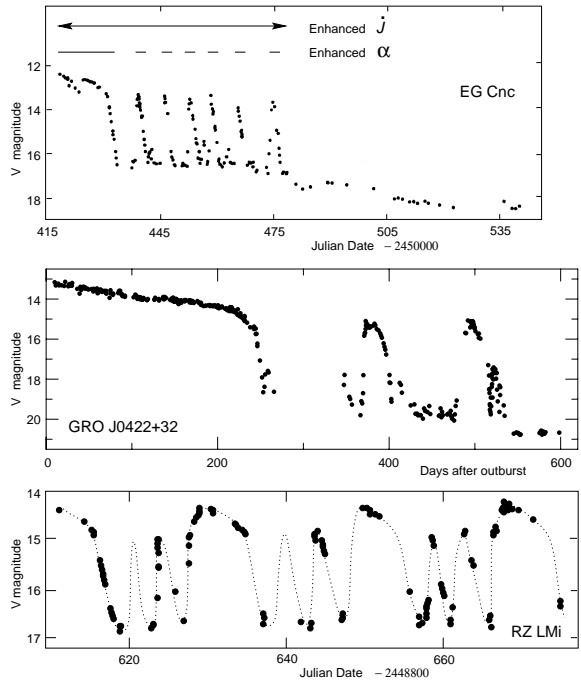}
\caption{The lightcurves of EG~Cnc ({\it top\/}) and the SXT GRO\,J0422+32 
({\it middle\/}) show echo outbursts after a superoutburst. The 
lightcurve of RZ~LMi ({\it bottom\/}) shows a 19-day supercycle.  
[Data from Patterson \etal\ 1998 (EG~Cnc), Kuulkers \etal\ 1996 
(GRO\,J0422+32) and Robertson \etal\ 1995 (RZ~LMi).]}
\end{figure}

\subsection{The problem of ER~UMa supercycles}
The ER~UMa stars have short supercycles lasting 40--50 days
(V1159~Ori and ER~UMa itself), 25--35 days (DI~UMa) and 19
days (RZ~LMi) (Fig.~1; Robertson, Honeycutt \&\ Turner 1995; 
Kato \etal\ 1999).

Osaki (1995b) has shown that, in the TTI model, increasing the
\mdot\ reduces the inter-outburst intervals, and thus
leads to a shorter supercycle. However, this effect can reduce the
supercycle only to $\approx$\,40 days (corresponding to 
\mdot\,$\approx 4 \times 10^{16}$ g s$^{-1}$);
increasing \mdot\ 
further results in longer superoutbursts, and thus the supercycle
lengthens again.  To achieve a supercycle of only 19 days,
the superoutburst must be truncated prematurely. Osaki (1995c)  
modelled RZ~LMi's supercycle by artificially ending the superoutburst
when the disk had shrunk from 0.46$a$ to only 0.42$a$, whereas an
end-value of 0.35$a$ is used for other SU~UMa stars.

A second problem with ER~UMa stars is that they show superhumps
beyond the end of the superoutburst, including in quiescence
(e.g.\ Patterson \etal\ 1995). This again violates the 
standard TTI model since the disk radius is then at a minimum
(0.26$a$ in the ER~UMa model of
Osaki 1995b, well inside the resonance zone at 0.46$a$), and so
should not show superhumps.

\section{A TTI variation for systems with ultra-low mass ratios}
In the standard TTI model, the
extra drain of angular momentum in an eccentric disk provides
a sufficient mass-flow to ensure that the disk remains in the
hot state, and thus the end of superoutburst requires the end
of the eccentricity. However, the above difficulties are solved
if the tidal and thermal instabilities are decoupled, allowing the
superoutburst to end when a disk that is still eccentric drops out
of the hot state.
This, in any case, is virtually required by the observations
of superhumps beyond the end of the superoutburst in ER~UMa and
WZ~Sge systems. 

I suggest that the key factor causing this behaviour is an
ultra-low mass ratio.  Fig.~3 shows the radius at which the
disk is truncated by tidal effects, and also the radius of
the 3:1 superhump resonance, both as a
function of mass ratio, $q$ (see Whitehurst \& King 1991).
For systems with $q$\,$\ga$\,0.3 the disk cannot extend to
the resonance radius, and the binary is not an SU~UMa star. 
As the mass ratio decreases, there is increasing room in the
disk outside the 3:1-resonance radius.
In this region, other, weaker, higher-order resonances operate,
and these eventually combine to produce the tidal limit.
However, the lower the mass ratio, the more likely that matter
can accumulate in regions of the disk outside 0.46$a$, which, avoiding
the main superhump resonance, interact less strongly with the
secondary star, and so suffer a lower tidal dissipation of
angular momentum, $\dot{J}$.

Another factor is that the strength of the tidal interaction 
decreases for lower mass ratios (assuming a fixed orbital period
and white-dwarf mass), and this might assist in decoupling the
tidal and thermal instabilities at ultra-low mass ratios.  
However, it should be noted that the tidal interaction 
is generally stronger at shorter orbital periods,
owing to the smaller size of the binary, which would counteract
this effect.

\begin{figure}[t]  % Fig 3
\plotone{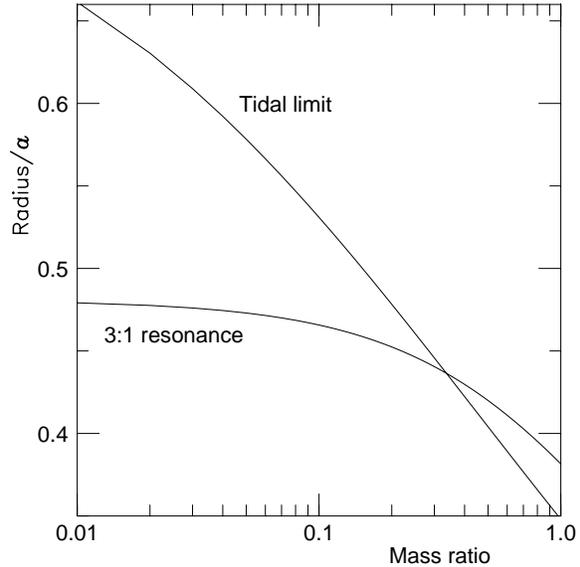}
\caption{The variation in the 3:1-resonance radius and the tidal-limit
radius (in units of the separation $a$), as a function of mass ratio.  
Based on Whitehurst \&\ King (1991).}
\end{figure}

In SU~UMa stars we can investigate $q$ without having
to measure masses, since $q$ correlates with the amount by
which the superhump period exceeds the orbital period, \leps,
defined by \leps\ = $(P_{\rm sh} - P_{\rm orb})/P_{\rm orb}$.
The exact relation between \leps\ and $q$ is not certain (see,
e.g., Patterson 1998; Murray 2000), but for current purposes
we require only that \leps\ increases monotonically with $q$.

Note that the stars giving difficulty to the TTI model have
exceptionally low values of \leps, and thus $q$ (Fig.~4).
EG~Cnc has the lowest value of all, followed by two other
stars that have shown echos (WZ~Sge and AL~Com).  DI~UMa is
one of two stars with supercycles shorter than 40 d, while 
\leps\ for the other, RZ~LMi, has not yet been measured.

\subsection{Application to echo outbursts}
Applying the idea to EG~Cnc, I suggest that the disk was larger
than 0.46$a$ and eccentric for the entire $\approx$\,75-d period
of superoutburst and echos (again, this is required by the
observation of superhumps thoughout this period by Patterson
\etal\ 1998). However, after 15 days of superoutburst the disk
outside 0.46$a$ was depleted sufficiently to revert to the cool
state --- the lower tidal $\dot{J}$ in this region could not
drive the disk edge inward quickly enough to prevent this. 
Thus a cooling wave, originating outside $0.46$a, ended the outburst.
However, the enhanced $\dot{J}$ from the 3:1 resonance
did ensure an enhanced flow into the inner disk, and this
triggered a new outburst, which originated in a heating wave from
the inner disk.  The fact that this outburst was short suggests
that the heating wave petered out before reaching the edge of
the disk, and thus a cooling wave quickly followed.
Thus a succession of short outbursts ensued, until the disk
had depleted sufficiently to revert to the circular state.
Then, when $\dot{J}$ declined, the mass-accretion rate dropped, and
as the last superhumps died away the system faded to full
quiescence.

Turning now to WZ~Sge and AL~Com, both of these systems, near the end of
outbursts in 1978 and 1995, respectively, showed a brief drop in magnitude 
followed by the resumption of the superoutburst (Patterson \etal\ 1981; 
Patterson \etal\ 1996). This again suggests that the
eccentric $\dot{J}$ was insufficient to sustain the outer disk
in the hot state, resulting in a cooling wave. But, in these
systems, the heating wave did run right to the edge of the disk,
and thus re-established the superoutburst. 
Note that the first cooling waves in WZ~Sge and AL~Com occurred 
25--30 days after the start of superoutburst,
whereas that in EG~Cnc occurred much earlier, at 15 days. 
Thus WZ~Sge and AL~Com appear to be systems on the borderline
of showing the echos that are fully fledged in EG~Cnc. 

\subsection{Application to ER~UMa supercycles}
As noted above, supercycles shorter than $\approx$\,40 days (those
of DI~UMa and RZ~LMi) 
require a premature end to the superoutburst.  As in EG~Cnc, I
suggest that a cooling wave from the lower-$\dot{J}$ region outside
0.46$a$ causes the disk to revert to the cold state while still
eccentric. The early end to the superoutburst means that the disk
retains a larger fraction of mass than is usual. This, and the
enhanced $\dot{J}$ from the 3:1-resonance region, means that
the flow of material into the inner disk soon triggers another
outburst. Again, the heating wave does not propogate to the outer
edge, and so the new outburst is short.

\begin{figure}[t]    %  Fig 4
\plotone{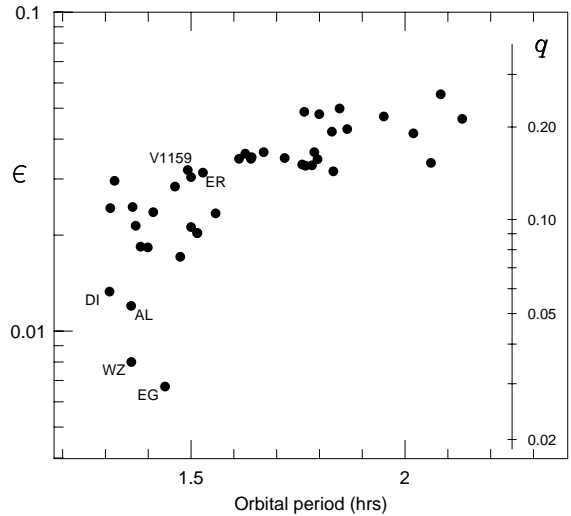}
\caption{The superhump period-excess, \leps, versus orbital period,
for SU~UMa stars. The data, and the translation to mass-ratio, $q$,
are taken from Patterson (1998).}
\end{figure}

Such stars therefore exhibit `echo' outbursts identical to
those in EG~Cnc --- short, but with superhumps.  The difference
is that in EG~Cnc the mass-transfer rate, \mdot, was much lower than the
mean mass-accretion rate, \maccdot, during the echos.  Thus the disk shrank
to quiescence.  In RZ~LMi and DI~UMa \mdot\ is
higher than \maccdot, and the disk gains material 
until a heating wave {\it does\/} propogate to the outer edge
of the disk, resulting in another superoutburst.

The two other ER~UMa stars, V1159~Ori and ER~UMa itself, have
longer supercycles (43 and 48 d) that can be explained in the
TTI model solely
by a high \mdot\  (Osaki 1995b). Further, their \leps\ values
imply mass ratios that are typical of SU~UMas (Fig.~4).
Thus, the TTI variant discussed above might be required only
for RZ~LMi and DI~UMa. However, the observation of superhumps
in V1159~Ori beyond the end of the superoutburst (Patterson \etal\
1995) suggests that the above ideas might come into play in all
ER~UMa stars. 

\subsection{Application to soft-X-ray transients}
Kuulkers \etal\ (1996) have pointed out the similarities between 
soft X-ray transients and WZ~Sge stars. SXTs are, for many 
intents and purposes, dwarf novae with a black hole replacing the white 
dwarf. The greater mass of the black hole in SXTs ($\sim$\,10\,M$_{\odot}$ 
compared to a white dwarf's $\sim$\,1\,M$_{\odot}$) ensures that they have 
low mass ratios. Thus it is notable that they can show echo outbursts 
(most obviously in GRO\,J0422+32, Fig.~2; Kuulkers \etal\ 1996) along with
quiescent superhumps around the time of the echo outbursts. From 
superhumps of GRO\,J0422+32, O'Donoghue \&\ Charles (1996) measure 
\leps\,=\,0.016, which is lower than in most SU~UMa stars (Fig.~4).
I propose that the echos arise for the same reason as they
do in EG~Cnc: the ultra-low mass ratio allows the disk to
cycle over the thermal instability while still in the eccentric state.
Note, though, that this requires that the greater irradiation of
an SXT does not prevent the outer disk from cooling. Perhaps the disk is 
convex, and the outer disk is shielded from the X-ray flux.

\section{Conclusions and predictions}
Osaki (1996) has outlined a unification model for cataclysmic
variables in which the different outburst properties are 
controlled by the two parameters $q$ and \mdot. I propose that 
echo outbursts and ultra-short supercycles ($\la$\,40 d) 
can be brought into the same scheme, and that they
occur in binaries with mass ratios of $q$\,$\la$\,0.07.
In such systems the superoutbursts end prematurely because
the lower tidal $\dot{J}$ outside the 3:1-resonance radius allows
cooling and heating waves to run through the disk while it is
in the eccentric state.  This produces `echo' outbursts
characterised by short durations and the presence of superhumps.
If \mdot\ $<$ \maccdot\ over the echo period, the system
subsides to quiescence (EG~Cnc behavior); if \mdot\ $>$
\maccdot, another superoutburst results (RZ~LMi behavior).

This model predicts that superhumps should be present (even in
quiescence) thoughout echo periods, and perpetually in stars with
ultra-short supercycles. Also, any system showing these
characteristics should have an exceptionally low value
of the superhump excess \leps\ and mass ratio $q$. Of the systems
dicussed in this paper, we know \leps\ for all except RZ~LMi, the
star with the shortest supercycle;  I predict that it has a
value of \leps\ even lower than the 0.013 found for DI~UMa. 

\acknowledgments
I thank Joe Patterson, Andrew King and Brian Warner for valuable
comments on this work.

\end{document}